\documentclass{elsart}

\usepackage{graphicx}

\usepackage{amssymb}

\begin{document}

\begin{frontmatter}



\title{Cosmic rays above the ankle from Z-bursts}


\author{Graciela Gelmini\thanksref{label1}},
\ead{gelmini@physics.ucla.edu}
\author{Gabriele Varieschi\thanksref{label2}}
\ead{gvariesc@lmu.edu}
\ead[url]{http://myweb.lmu.edu/gvarieschi/}

\address[label1]{Dept. of Physics and Astronomy, UCLA (University of California, Los Angeles), 405 Hilgard Ave., Los Angeles, CA 90095-1547, USA}
\address[label2]{Dept. of Physics, Loyola Marymount University, One LMU Dr., Los Angeles, CA 90045-2659, USA}

\begin{abstract}
Neutrinos from far away sources annihilating at the Z resonance on relic neutrinos  may give 
origin to the ultrahigh-energy cosmic rays.
Here we present predictions of this mechanism with relic neutrinos lighter
 than 1 eV, which do not gravitationally cluster. We show that  not only the super GZK events, but the ``ankle" and all  events above it
can be accounted for. Most primaries above the ankle
are predicted to be  nucleons up to $10^{20.0}$~eV and  photons at
higher energies. We also find an accumulation at the GZK cutoff energy, a hint
of  which can  be seen in the data.
\end{abstract}

\end{frontmatter}

\section{Introduction}
\label{sec-1}

The existence of ultrahigh-energy cosmic rays (UHECR) with energies above the 
Greisen-Zatsepin-Kuzmin (GZK) cutoff~\cite{gzk}  of about $5\times 10^{19}$ eV, 
presents an outstanding problem~\cite{data}.  Nucleons and photons with those 
energies have short attenuation lengths and could only come from distances of 
100 Mpc or less~\cite{50Mpc,40Mpc}, while plausible
astrophysical sources for those energetic particles are much farther away. 

An elegant and economical solution  to this problem, proposed by T. 
Weiler~\cite{weiler}, consists of the production of the necessary photons and 
nucleons close to Earth, 
in the annihilation at the $Z$-resonance of  ultrahigh-energy neutrinos 
 coming from remote sources, $\nu_{\rm UHE}$, and relic background neutrinos.  
These events were named ``$Z$-bursts"  by T. Weiler.
One of the most appealing features of the ``$Z$-bursts" scenarios is that the 
energy scale of $10^{20-21}$~eV, at which the
unexpected events have been detected, is generated naturally given the possible
mass range of relic neutrinos. 
The $Z$-resonance occurs when the energy of the incoming $\nu_{\rm UHE}$ 
 is $E_{\nu_{\rm UHE}}= E_{Res}$, 
\begin{equation}
E_{Res} = \frac{M_Z^2}{2~m_\nu} ,
\label{Eres}
\end{equation}
where $m_\nu$ is the mass of the relic neutrinos. This is the new cutoff of the
UHECR energy in these models. It depends inversely on the mass of the relic 
neutrinos.  Since arguments of structure formation in the Universe show  $m_\nu <$ few eV,
 then $E_{\nu_{\rm UHE}} > 10^{21}$ eV, precisely above the GZK 
cutoff, as needed. 

In this paper we concentrate on a particular ``Z-bursts'' 
scenario~\cite{gk1,gk2},
in which the relic neutrinos are lighter than 1 eV.
These lighter neutrinos, contrary to those in the original scenario, cannot be 
gravitationally bound, they have a constant density over the whole Universe. In 
particular, we concentrate on relic neutrinos with mass compatible with 
Super-Kamiokande results, if neutrino masses are hierarchical (however the results
we obtain apply to heavier relic neutrinos, while light enough to not cluster).
Super-Kamiokande has provided a strong evidence for the oscillation in 
atmospheric showers of  two neutrino species with masses $m_1$,
$m_2$ and  $\delta m^2 $ = $m_1^2-m_2^2$ = $(1-8) \times 10^{-3}$ eV, 
consisting
mostly of about equal amounts of $\nu_\mu$ and another flavor eigenstate 
neutrino, $\nu_\tau$ or a sterile neutrino~\cite{SK}.  If
neutrino masses are hierarchical, as those of the other leptons and quarks,
then the heavier of the two oscillating neutrinos, call it $\nu_{\rm SK}$, has 
a mass $m_{\rm SK} =\sqrt{\delta m^2} \simeq 0.07$ eV.
With $m_\nu = m_{\rm SK}$, the new UHECR cutoff becomes 
\begin{equation}
E_{Res} \simeq 0.6 \times 10^{23}~eV .
\label{Eres2}
\end{equation}
Due to the large multiplicity of the Z-decays, after energy losses in the 
propagation (as shown in detail here) this value of $E_{Res}$ predicts 
many super GZK UHECR events (many more than with eV relic neutrino 
masses). 

We agree with Farrar and Piran~\cite{farrarpiran}, who have argued 
that any mechanism accounting for the events beyond the GZK cutoff  
should also account for the events down to the ankle, including their isotropy 
and spectral smoothness. We show here that the  model we consider can account 
for the ankle and all the events above it if 
the  position of the ankle is close to that measured by AGASA, $E_{ankle} = 
10^{19.0}$~eV (see \cite{naganowatson}, in particular Table V, 
and references therein). Moreover, a  reliable prediction 
of the model is that most primaries above the ankle should 
be nucleons  up to about $10^{20.0}$~eV and  
photons at higher energies. We also find that  nucleons do accumulate at the 
GZK cutoff energy, which could 
account for the hint of a slight accumulation seen in the data. Photons become 
dominant at energies higher than  $10^{20}$~eV in our model. So
these photons are all above the threshold energy (which is about $ 5 \times 
10^{19}$~eV) 
to pair produce in the Earth's magnetic field (which should generate a certain 
amount of north-south asymmetry in the arrival direction distribution). 

Let us return to the issue of the isotropy of the events above the ankle, i.e., the absence of a correlation with the galactic halo. Because the relic neutrinos we assume do not gravitationally cluster, the isotropy of the events reflects the isotropy of the ultrahigh-energy neutrino sources.
 In particular, relic neutrinos of mass $m_{\rm SK}$ require a large flux of neutrinos with energies of the order of $10^{23}$~eV.
It is unlikely that active galactic
nuclei~\cite{wb}, neutron stars~\cite{olinto1}, or other astrophysical
sources could produce such a high energy flux of ultrahigh-energy neutrinos.
Topological defects~\cite{berez}, or superheavy relic
particles~\cite{gk2}, could instead easily generate the requisite flux of primary
neutrinos (but there are still problems with these sources).
For example, with unstable superheavy relic particles, 
which form part of the cold dark matter, decaying
mostly into neutrinos~\cite{gk2},  the directions 
of UHECR could map the distribution of parent particles (which should coincide 
with the distribution of cold dark matter) at large redshifts.  This is 
because the initial energy of the $\nu_{\rm UHE}$ produced in the decay needs 
to be redshifted to the energy of the ``$Z$-burst" in its way to the Earth. 
 Directional clustering, evident in the existing
data~\cite{clustering}, would then reflect the distribution of
matter at a certain red shift determined by this process of ``cosmological
filtering''. Thus, absence of directional correlations with the galactic halo, 
as well as  directional clustering, could be easily accommodated~\cite{gk2}.

Besides, $\nu_{\rm UHE}$ produced by  unstable superheavy relic particles would 
have a spectrum opposite to an astrophysical spectrum, growing rapidly with 
energy, up to a sharp cutoff at an energy of the order
of the parent particle mass. This spectrum has almost no neutrinos at low 
energy where bounds exist~\cite{gk2,sigl}.  
Most bounds on  ``$Z$-bursts" models (see for example \cite{wb}) assume 
that the $\nu_{\rm UHE}$ have a typical astrophysical spectrum, decreasing with
energy as $E^{-\gamma}$, with $\gamma$ a number of order one.  These bounds do
not hold if the $\nu_{\rm UHE}$ spectrum has a very different energy 
dependence. However, a model for these parent particles is arguably 
difficult to obtain~\cite{CDF,Uehara}. Moreover,
the EGRET bound on the diffuse low-energy gamma ray flux
 resulting from the ``$Z$-bursts" imposes important constraints \cite{EGRET}, which might rule out heavy particles decaying mostly into neutrinos as sources \cite{Berezinsky}.  

In the next section we present our simulations and the resulting spectrum of UHECR.
We would like to point out that the main result of this paper, the fact that
``$Z$-bursts" can account for the ankle and all events above it, does also hold for larger relic neutrino masses, for which the problem of the sources becomes less severe. In fact
even if we used $m_{\rm SK}$  here, our considerations apply with trivial changes to other  
masses for which relic neutrinos are too light to gravitationally cluster. As the relic neutrino mass increases, all the features in the spectrum we find here should move progressively to lower energies.

\section{Spectrum of UHECR from ``$Z$-bursts"}

We have performed simulations of the photon, nucleon and 
neutrino fluxes coming from a uniform distribution of ``$Z$-bursts",
namely  $\nu \bar{\nu}$ annihilations at the Z pole ($\nu
\bar{\nu} \rightarrow Z \rightarrow p \gamma ...$),
with the energy of Eq. 2, corresponding  to relic 
neutrinos of mass $m_{\rm SK}$.  The ``$Z$-bursts" were simulated using PYTHIA 
6.125 \cite{PYTHIA}, and the absorption of photons and nucleons was modeled 
using energy attenuation
lengths provided by Bhattacharjee and Sigl~\cite{BS}, supplemented by 
radio-background models by Protheroe and Biermann~\cite{PB}. 

We simulated a uniform distribution of about $10^7$ Z
particles up to a 
maximum $z_{max}=2$. Even if the shape of the spectrum of
nucleons and photons changes somewhat with other Z particle distributions,
the main features of the spectrum stated here remain the same.

The decay of the Z bosons through all possible channels was done automatically
by the PYTHIA routines~\cite{PYTHIA}, using the default options of this
program. For comparison with the other figures we show in Fig.~1 the spectra 
given by PYTHIA, normalized per Z boson (not including redshifts and energy 
absorption).  The multiplicities that PYTHIA gives  per Z-decay are (in each case 
counting particles and antiparticles) 1.6 nucleons, 17 photons, 15 
$\nu_e$, 30 $\nu_{\mu}$ and 0.23 $\nu_{\tau}$.


\begin{figure*}
\hspace*{-20.0mm}
\includegraphics{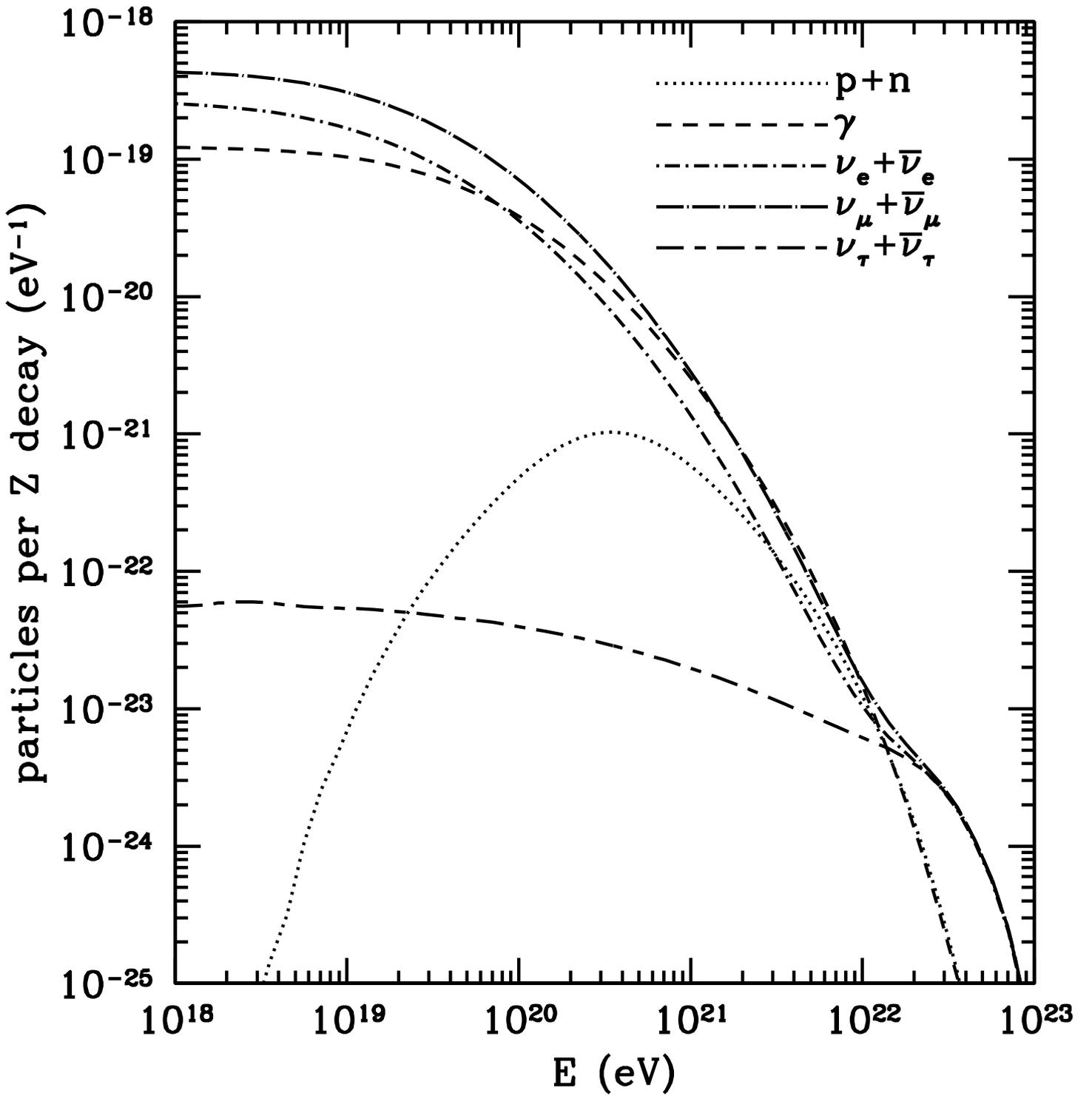}
\caption{\label{fig1}Spectra  of stable particles produced per Z decay by PYTHIA (no absorption or redshift included).}
\end{figure*}

In our simulation, each Z boson generated by PYTHIA was placed on the ``event 
list'' of the cascade generator at a randomly selected distance.  The cascade 
of decay products was then boosted. The 
$\gamma$ factor corresponding to an energy  $E_\nu=E_{res}$ is
\begin{equation} 
\gamma={\frac{E_\nu+m_\nu}{M_z}}= 6.25 \times 10^{11}.   
\label{eq:sim3} 
\end{equation}  
The gamma factor of each boost was  actually corrected to include the redshift 
of the decaying  Z particle. We then followed the propagation of the nucleons, photons and neutrinos 
resulting from the boosted Z decays. Neutrinos  do not interact in their 
propagation. Thus, the energy spectra of
the three kinds of neutrinos were simply generated by counting the number of
particles  per energy bin and normalizing this number to the total number of Z 
particles used. 

We included energy absorption for nucleons and photons. Each nucleon or
photon was created by PYTHIA in the
initial cascade at a fixed position, with  fixed energy and direction of
motion with respect to the Earth frame of reference.
The distance  each particle had to travel before reaching  Earth was
 compared with the appropriate attenuation length in space for the
particle energy.  If the distance  was smaller than the
attenuation length, the particle was assumed to  reach  Earth unchanged.
 In the opposite case, the energy and
momentum of the particle were degraded by a factor $1/e$ after traveling
a distance equal to the attenuation length (and the particle was
placed again in the list constituting the cascade at the new 
position,  with the degraded energy and momentum). 

This process was continued
until all nucleons and photons reached Earth and were counted
in the final spectra, or until their energies became too small to be
significant, in which case they were simply discarded from the cascade.
At this point, the final nucleon and photon spectra were 
normalized to the total number of Z particles used (the same  done with the 
neutrino spectra). The results are given in Fig. 2, with an approximate fit to 
the AGASA cosmic ray data \cite{AGASA}.
\begin{figure*}
\hspace*{-20.0mm}
\includegraphics{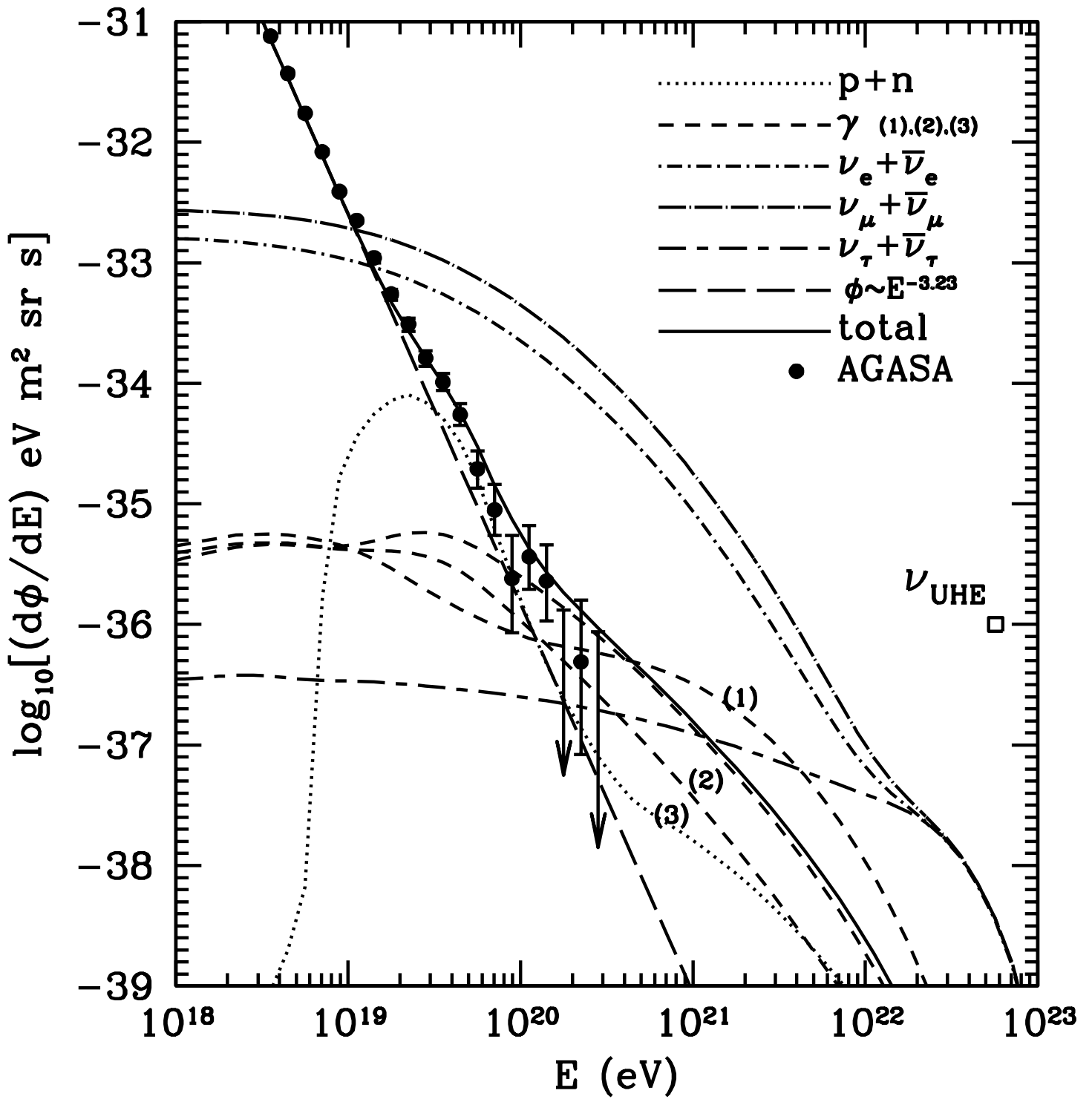}%
\caption{\label{fig2}Predicted spectra  from ``$Z$-bursts" with a uniform distribution up to $z=2$, added to a power law spectrum which fits the 
data below the ankle. Primaries above the ankle are nucleons up to $10^{20.0}$~eV and  photons at higher energies.}
\end{figure*}
We used the nucleon energy attenuation length given by Bhattacharjee and Sigl
in the  Fig.~9 of Ref. \cite{{BS}}, 
based on results from Ref. \cite{stecker} and \cite{halzen}. 

The energy attenuation length of photons is poorly known, due to the 
uncertainty  in the radio background. Using the  attenuation length shown in 
Fig.~11  of Ref. \cite{BS},  based on  observations of 
Clark et al.~\cite{clark}, the resulting
photon flux is shown in Figs.~2 and~3 as curve~(1).  Protheroe and 
Biermann~\cite{PB} produced two models for the radio background
which lead  to shorter interaction lengths than those based on Clark et al. 
 From the provided interaction lengths we constructed approximate 
attenuation lengths for the models of Protheroe and Biermann by
reducing the  attenuation length based on Clark et al. by the difference 
between the interaction lengths. With the attenuation lengths so constructed we
obtained the curves (2) and (3) in Figs.~2 and~3. This is obviously only an 
approximation, since the mean interaction and energy attenuation lengths do not 
have exactly the same energy dependence. However,  we believe the three curves 
we obtained give a good representation of the possible range of predicted 
photon fluxes, in view of the uncertainties related to the energy attenuation 
of photons in space.

We have arbitrarily used the middle photon flux, curve number (2), when 
computing the total flux. We added the proton and photon fluxes obtained in 
this paper to a power law spectrum with slope $-3.23$ found by AGASA to fit the 
data below the ankle (from $10^{17.6}$ to $10^{19.0}$ eV; see Table V of 
\cite{naganowatson}). Our results  depend very little on which  of the three 
photon fluxes we use.

The fit of the AGASA data with our total flux provides the normalization of 
the photon and nucleon differential fluxes $F$, denoted  as $d\phi/dE$ in the 
figures (in this case $F_{AGASA}$ = $10^{-14.2}~({\rm m^2~sr~s})^{-1}$ 
$F_{PYTHIA}$ = $ 6 \times 10^{-15}~({\rm m^2~sr~s})^{-1}$
$F_{PYTHIA}$), that allows us to determine the (assumed homogeneous
and isotropic) flux of ultrahigh-energy neutrinos close to the $Z$-resonance 
energy of Eq. 2 
(at some energy between $E_{Res}/ (1 + z_{max})= E_{Res}/3$ 
and $E_{Res}$) to be 
\begin{equation}
F_{\nu_{\rm UHE}} \simeq 1\times 10^{-36} \frac{1}{{\rm eV~ m}^2 {\rm ~sr~ s}}~~ ,
\label{Fnu}
\end{equation}
if no lepton asymmetry is assumed in the neutrino background. This flux is 
shown in Figs.~2 and 3, with the label ``$\nu_{\rm UHE}$". With the level of accuracy of
our simulation we can only determine the order of magnitude of this flux. Moreover,
 for the light neutrinos we are considering, 
present experimental bounds allow for a  lepton asymmetry which could 
increase the number of relic neutrinos by  a factor smaller than 10, reducing 
the necessary $F_{\nu_{\rm UHE}}$ by the same factor. 
 
 We have continued the power law spectrum  accounting for the events below the 
ankle, presumably due to galactic sources, way beyond the ankle, while
this contribution may die out at or soon above the ankle. In any event,
without adding up the power law, i.e., taking into account only 
the photon and nucleon fluxes from ``$Z$-bursts" computed here, we had made a 
very similar fit~\cite{gelminitalk}, with which we had obtained the same 
$F_{\nu_{\rm UHE}}$.

As we just mentioned, without making an assumption on the spectrum of $\nu_{\rm UHE}$, the 
``$Z$-bursts"  mechanism provides an estimate of the $\nu_{\rm UHE}$ flux only close to  
the  $Z$-resonance energy $E_{Res}$ (at some energy between
$E_{Res}/ (1 + z_{max})= E_{Res}/3$
and $E_{Res}$) .  It is interesting to point out that,  
on the  basis of that one estimate,  the ``Goldstone 
Experiment"(GLUE)~\cite{GLN} could start testing our
model after about 300 hours of observation. This
 can be seen in Fig. 3, where we show the present limits of this experiment, 
based on about 30 hours of observation. In Fig.~3 the fluxes of Fig. 2 have 
been multiplied by $E^2$. GLUE searches for lunar radio emissions from
interactions of neutrinos above $10^{19}$ eV of energy and is 
expected to accumulate 120 hours of observation later this year.

Figs. 3 and 4 include Akeno data taken from Fig. 23 of
Ref.~\cite{naganowatson}, originally from Ref.~\cite{Akeno} (besides the AGASA 
data). In Fig.~4 we show our nucleon flux, only our middle  photon flux (curve 
(2) in previous figures), the power law providing a good fit to data below the 
ankle, and the total flux (the sum of these three components).


\begin{figure*}
\hspace*{-20.0mm}
\includegraphics{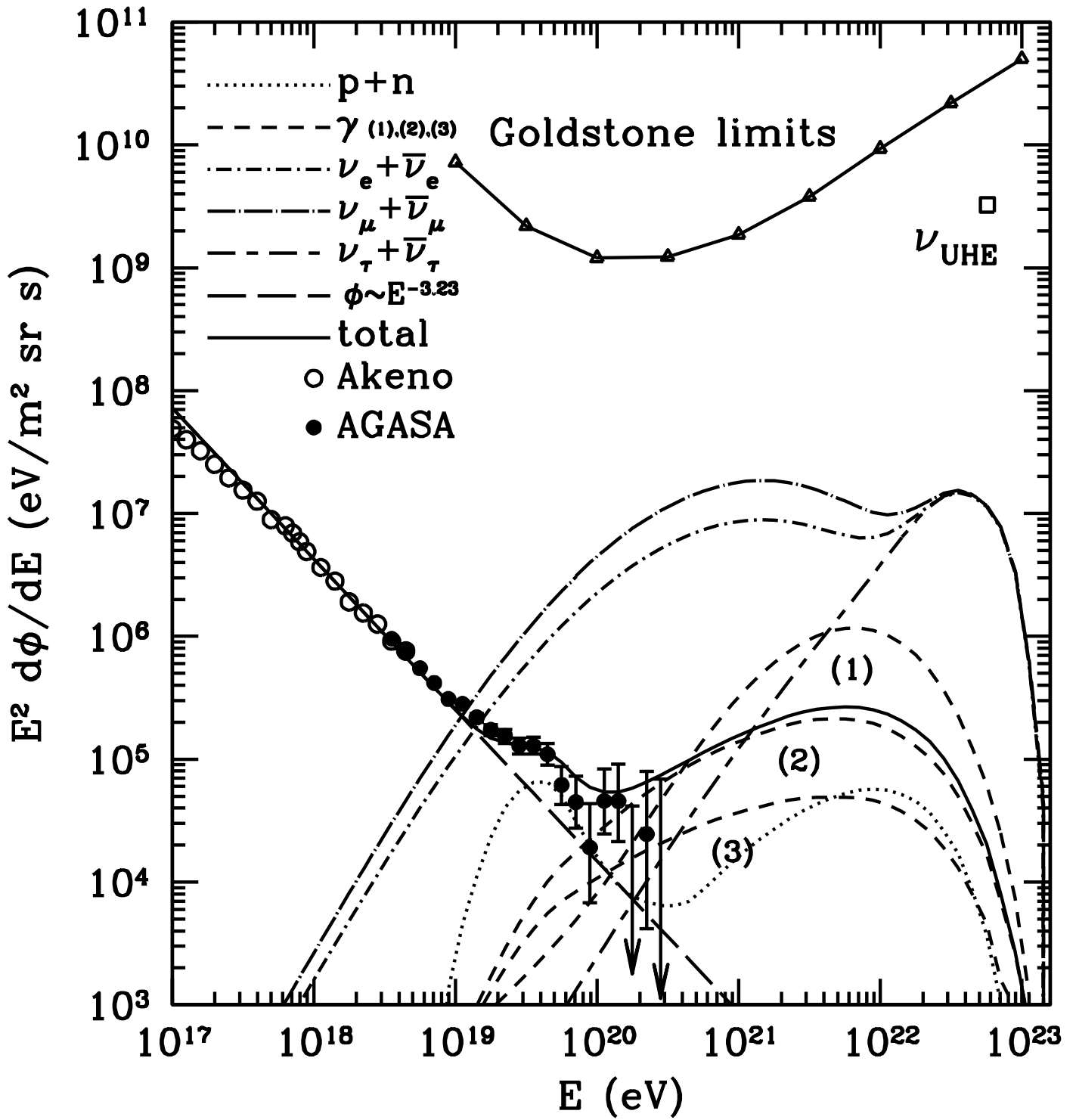}%
\caption{\label{fig3}The fluxes of Fig.~2 have been multiplied by 
E$^{2}$. Data from Akeno are shown  besides the AGASA data. ``$\nu_{\rm UHE}$"
labels the flux of UHE neutrinos predicted close to the Z-resonance energy.}
\end{figure*}

We believe that Fig.~4, in which the plotted spectra have been multiplied by
$E^3$, shows best the change of slope close to the  
position of the ankle as measured by AGASA, $E_{ankle} = 10^{19.0}$~eV.
 Obtaining a value of $E_{ankle}$  a  factor of 3 smaller, close to 
$10^{18.5}$~eV as 
measured by Fly's Eye (see Table V of \cite{naganowatson}) may require to 
increase the $z_{max}$ of the source distribution considerably
(which would point to particles as sources), since this would move the lower 
energy edge of the nucleon flux to lower energies. Alternatively, larger values 
of the relic neutrino mass may also work to lower $E_{ankle}$, since all the 
features in the spectrum should move to lower energies (even if we used
 $m_{\rm SK}$  here, our considerations apply with trivial changes to other  
masses for which relic neutrinos are too light to gravitationally cluster).

 In Fig.~4 one can clearly see the enhancement of 
the predicted flux at the GZK cutoff energy, at about $5 
\times 10^{19}$~eV, due to the accumulation of nucleons,  which can also be 
seen in the AGASA data.


\begin{figure*}
\hspace*{-20.0mm}
\includegraphics{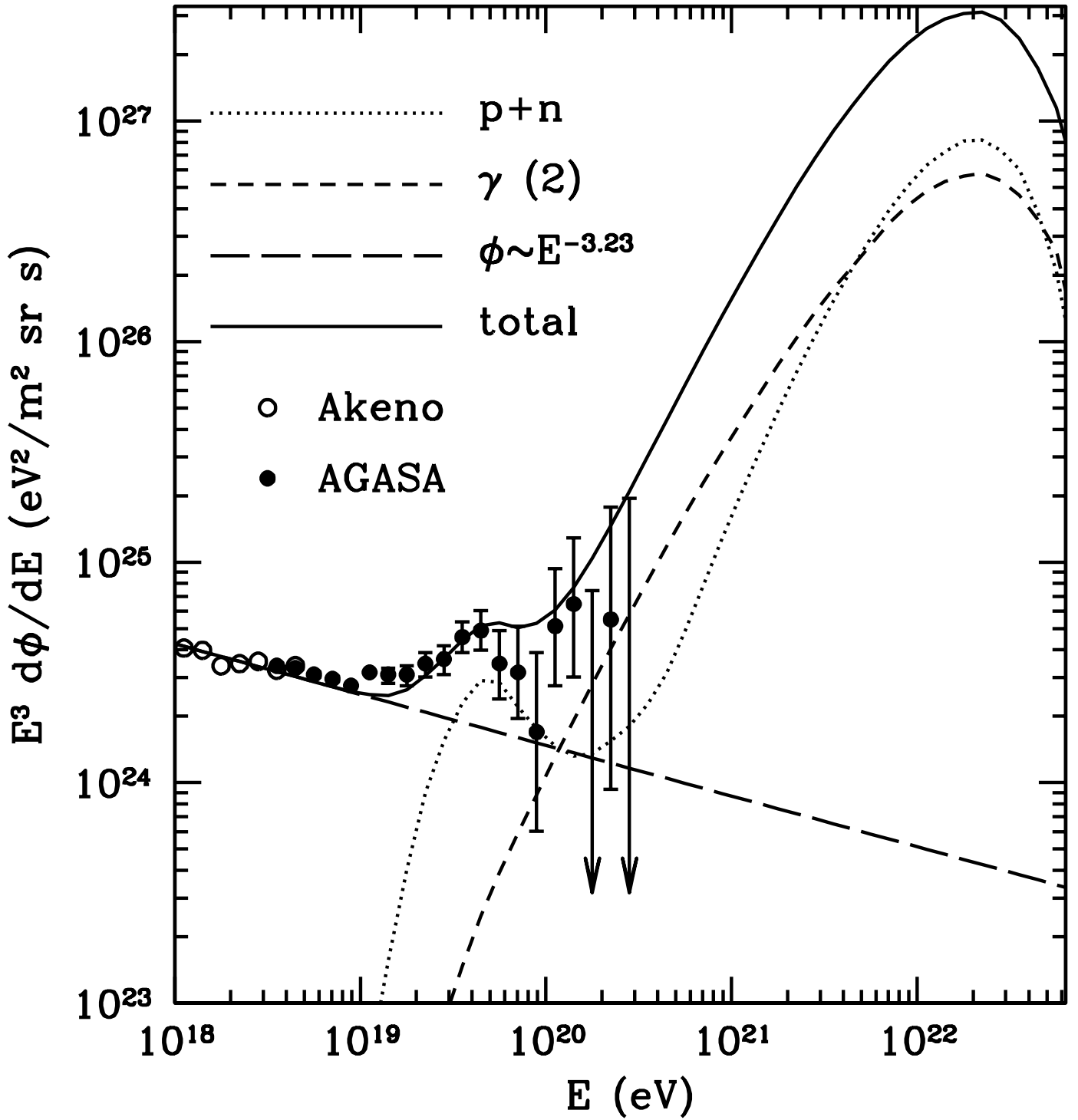}%
\caption{\label{fig4}The fluxes of Fig.~2 have been multiplied by 
E$^{3}$. The  position of the ``ankle" is close to  $10^{19.0}$~eV as measured 
by AGASA. The predicted  accumulation at the GZK cutoff energy, about $5 
\times 10^{19}$~eV,  can also be seen in the data.}
\end{figure*}

\section{Conclusions} 

In this paper we considered  a particular ``Z-bursts'' 
scenario~\cite{gk1,gk2}, in which the relic neutrinos are lighter than 1 eV, and thus
do not gravitationally cluster. Using in particular 0.07 eV relic neutrinos, 
we have shown that ``$Z$-bursts"  may account  not 
only for the super GZK events, but for the ``ankle" and all UHECR events above it, 
including their isotropy and spectral smoothness. In our simulation we found 
  the ``ankle"  close to $E_{ankle} = 10^{19.0}$~eV as measured by AGASA. 
A reliable prediction of the model is that most primaries above the ankle 
should be nucleons up to about $10^{20.0}$~eV and  
photons at higher energies. Moreover, the nucleons do accumulate at the GZK 
cutoff energy, which could  account for the hint of a slight accumulation seen 
in the data. The model predicts a new cutoff, which with 0.07 eV relic neutrinos 
is at $E_{Res} \simeq 0.6 \times 10^{23}~eV$.

We have not included the effect of extragalactic magnetic fields, thus for the 
predictions of this paper to be true these fields should be sufficiently small, 
probably  smaller than $10^{-9}~G$.

Finally let us comment on recent related papers. Photon and nucleon spectra very similar to those presented here are shown
in Fig. 2a of Ref. \cite{kalashev} corresponding to ``$Z$-bursts" with 0.1 eV 
relic neutrinos and a different model for the  distribution of ``$Z$-bursts"
with redshift up to $z_{max}= 3$. This model satisfies the EGRET bound on low 
energy photons (even if with astrophysical sources emitting only UHE 
neutrinos).  This reassures us that the result we present here is robust. 
The EGRET bound has been computed for various ``$Z$-bursts"
scenarios~\cite{kalashev,Yoshida}, including the particular one we 
concentrated on here~\cite{EGRET}, which seems to work well with sources emitting only
UHE neutrinos. The most serious problem with these sources may be the electroweak cascading of the UHE neutrinos produced in the decays, as recently claimed in~\cite{Berezinsky}.  

 Events above the ankle have been previously fitted with ``$Z$-bursts" products plus an 
additional hypothetical component of galactic or extragalactic protons in 
Ref.~\cite{Fodor},
varying the relic neutrino mass, in an attempt to provide a 
determination  of this mass using UHECR data. However the normalization and 
slope of the extra proton flux and the normalization of the nucleon flux from 
``$Z$-bursts" (photons from ``$Z$-bursts" were neglected) were taken as 
fitting parameters. This procedure does not make clear if it is actually the ``$Z$-bursts" 
which account for the change of slope at the ``ankle" and for the events above the 
``ankle". Moreover, with this procedure~\cite{Fodor}, there is no prediction  of  
the position of the ankle, since this is one of the parameters to be fitted 
by the sum of the mentioned two fluxes. Here we took instead the measured flux below the ankle, with the measured slope and normalization. We considered it
to be of galactic origin, as the correlation with the galactic center
of the arrival directions measured by AGASA around $10^{18}~eV$~\cite{correlation} seems to indicate. Then we added to this fixed flux the new component due to ``$Z$-bursts", which led us to find a prediction for the position of the ankle, and the 
normalization of the flux of primaries due to ``$Z$-bursts". 

We believe our flavor of ``$Z$-bursts" provides  a plausible  
explanation to the puzzle of ultrahigh-energy cosmic rays, not 
only for the super GZK events, but for the ``ankle" and all  events above it.

\vspace{5 mm}


This work was supported in part by the US Department of Energy grant
DE-FG03-91ER40662, Task C. 
G.V. was also supported by an award from Research Corporation. 
We thank A. Kusenko and S. Nussinov for many 
valuable discussions, and P. Biermann and E. Roulet for comments and
 suggestions.



\end{document}